\documentclass[preprint,showpacs,preprintnumbers,amsmath,amssymb,eqsecnum]{revtex4}

\usepackage{graphicx}
\usepackage{dcolumn}
\usepackage{bm}
\usepackage{verbatim}

\def\z{{\mathbf{z}}}

\begin{document}


\title{Quantum Arrival Time For Open Systems}

\author{J.M.Yearsley}
\email{james.yearsley@imperial.ac.uk}

\affiliation{Blackett Laboratory \\ Imperial College \\ London SW7
2BZ \\ UK }

\begin{abstract}
We extend previous work on the arrival time problem in quantum mechanics, in the framework of decoherent histories, to the case of a particle coupled to an environment. The usual arrival time probabilities are related to the probability current, so we explore the properties of the current for general open systems that can be written in terms of a master equation of Lindblad form. We specialise to the case of quantum Brownian motion, and show that after a time of order the localisation time the current becomes positive. We show that the arrival time probabilities can then be written in terms of a POVM, which we compute. We perform a decoherent histories analysis including the effects of the environment and show that time of arrival probabilities are decoherent for a generic state after a time much greater than the localisation time, but that there is a fundamental limitation on the accuracy, $\delta t$, with which they can be specified which obeys $E\delta t>>\hbar$. We confirm that the arrival time probabilities computed in this way agree with those computed via the current, provided there is decoherence. We thus find that the decoherent histories formulation of quantum mechanics provides a consistent explanation for the emergence of the probability current as the classical arrival time distribution, and a systematic rule for deciding when probabilities may be assigned.
\end{abstract}



\maketitle

\newcommand\beq{\begin{equation}}
\newcommand\eeq{\end{equation}}
\newcommand\bea{\begin{eqnarray}}
\newcommand\eea{\end{eqnarray}}

\def\A{{\cal A}}
\def\D{\Delta}
\def\H{{\cal H}}
\def\E{{\cal E}}
\def\D{{\mathcal{D}}}
\def\p{\partial}
\def\la{\langle}
\def\ra{\rangle}
\def\ria{\rightarrow}
\def\Z{{\bf z}}
\def\t{{\tau}}
\def\y{{\bf y}}
\def\k{{\bf k}}
\def\q{{\bf q}}
\def\p{{\bf p}}
\def\P{{\bf P}}
\def\r{{\bf r}}
\def\d{{\partial}}
\def\s{{\sigma}}
\def\a{\alpha}
\def\b{\beta}
\def\e{\epsilon}
\def\z{\xi}
\def\U{\Upsilon}
\def\g{{\gamma}}
\def\G{\Gamma}
\def\om{{\omega}}
\def\Tr{{\rm Tr}}
\def\iff{{\rm iff}}
\def\ih{{ \frac {i} { \hbar} }}
\def\trho{{\rho}}
\newcommand\bra[1]{\left<#1\right|}
\newcommand\ket[1]{\left|#1\right>}
\newcommand\brak[2]{\left<#1|#2\right>}
\newcommand\dif[2]{\frac{\partial #1}{\partial #2}}
\newcommand\diff[2]{\frac{\partial^{2} #1}{\partial #2^{2}}}

\def\au{{\underline \alpha}}
\def\bu{{\underline \beta}}
\def\pp{{\prime\prime}}
\def\id{{1 \!\! 1 }}
\def\half{\frac {1} {2}}

\def\jmy{james.yearsley@imperial.ac.uk}

\section{Introduction}

Questions involving time in quantum theory have a rich history, and there is still much debate about their status \cite{time}. Quantities such as arrival and dwell times, despite being measureable \cite{meas}, still lack concrete grounding within ``standard'' quantum theory, and there has been considerable interest in understanding  these quantities in the framework of various interpretations of quantum theory. Arrival times, in particular, have attracted much interest, as the natural procedure of quantising the appropriate classical quantity gives rise to an operator which is not self-adjoint and thus, in standard quantum theory at least, cannot easily be considered as an observable. 

Despite these difficulties, if one considers a free particle
in an initial state $\rho= \ket{\psi}\bra{\psi} $
localized in $x>0$ consisting entirely of negative momenta, and asks
for the probability $p(t_1,t_2)$ that the particle crosses the origin
during the time interval $[t_1,t_2]$ there {\it is} a semi-classical answer, given by the difference between the probability of being in $x>0$ at the initial and final time \cite{time}. Defining $P(t)=\theta(\hat x(t))$,
\bea
p (t_1, t_2)&=&\Tr(P(t_{1})\rho)-\Tr(P(t_{2})\rho)=-\int_{t_1}^{t_2} dt \Tr(P \dot \rho_{t})\nonumber\\
&=&\int_{t_{1}}^{t_{2}}dt J(t)
\label{1},\eea
where
\bea
J(t)&=&\frac{i\hbar}{2m}\left(\psi^{*}(0,t)\frac{\partial \psi(0,t)}{\partial x}-\frac{\partial \psi^{*}(0,t)}{\partial x}\psi(0,t)\right)\nonumber
\eea
is the standard quantum mechanical probability current. (We denote the state at time $t$ by $\rho_{t}$.) This can also be written in the following two forms,
\bea
p(t_{1},t_{2})&=& \int_{t_{1}}^{t_{2}}dt \int dp dq \frac{(-p)\delta(q)}{m}W_{t}(p,q)\label{1w},
\eea
where $W_{t}(p,q)$ is the Wigner function, defined later, and
\bea
p(t_{1},t_{2}) &=& \Tr(C\rho)\nonumber\\
C &=& \int_{t_1}^{t_2} dt  \frac { (-1) } {2 m } \left( \hat p \delta ( \hat x(t) ) + \delta (\hat x(t) ) \hat p \right)
\nonumber \\
&=& P ( t_1 ) - P (t_2 ).
\label{2}
\eea
These expressions agree with the classical result, with the Wigner function $W$ replaced by the classical distribution function $w$, provided that the classical trajectories are straight lines. They are also correctly normalized to 1
as $t_2 \rightarrow \infty $ with $t_1 = 0$, but they are not generally positive, even for states consisting entirely of negative momenta. This genuinely quantum phenomenon is called backflow
and arises because the operator $C$, positive classically, has negative
eigenvalues  \cite{cur,BrMe,back}. This means we cannot generally regard Eq.(\ref{1}) as providing an acceptable arrival time distribution. There is, in addition, a more fundamental problem with Eq.(\ref{1}), which is that probabilities in quantum theory should be expressible in the form \cite{NC}
\beq
p(\a)=\Tr(P_{\a}\rho).
\eeq
Here $P_{\a}$ is a projection operator, or more generally a POVM, associated with the outcome $\a$. Eq.(\ref{1}) cannot be expressed in this form, and we therefore conclude that it is not a fundamental expression in quantum theory, so must be the result of some approximation.

An interesting clue as to how to improve on Eq.(\ref{1}) is provided by the expression for the current in terms of the Wigner function, Eq.(\ref{1w}). Negativity of the Wigner function is a neccessary condition for the negativity of the current (in the sense of backflow, as discussed in \cite{HaYe1}), but it is known that evolution in the presence of an environment typically renders the Wigner function positive after a short time \cite{DoHa}. This suggests that something like Eq.(\ref{1}) may be an acceptable, if still heuristic, arrival time distribution for a system coupled to an environment, at least after some time. The first aim of this paper is to derive the correct analogue of Eq.(\ref{1}) for a system  coupled to an environment, and to prove that it can indeed be regarded as an arrival time distribution after a short time. In particular we show that the arrival time probabilities computed in this way may be written as the trace of the density matrix times a POVM.

However although this may allow us to interpret the current as the arrival time distribution, we are still left with the task of deriving Eq.(\ref{1}) from some more fundamental quantity. This has previously been achieved, in some approximation, in Refs.\cite{HaYe1, HaYe2} in the context of the decoherent histories approach to quantum theory \cite{GeH1,Gri,Omn,Hal2,DoH}, for the case of a free particle, and the derivation was shown to hold for states and intervals exhibiting sufficient decoherence. In decoherent histories the probability of arriving in an interval $[t_{1},t_{2}]$ is approximately given by
\beq 
p(t_{1},t_{2})=\Tr(C\rho C^{\dagger}),
\eeq
with $C$ given by Eq.(\ref{2}), and this reduces to Eq.(\ref{1}) under the somewhat special condition of decoherence. However it was also shown that there exist states for which decoherence of histories cannot be obtained for reasonable levels of coarse graining, and for which therefore we cannot assign a time of arrival probability distribution. States exhibiting backflow are good examples of this. 

Lack of decoherence for a general state is a situation frequently encountered in the literature on decoherent histories. The solution lies in the observation that realistic systems are always coupled to an environment, and that as such they are most fundamentally described by open system dynamics. Because by definition an environment consists of degrees of freedom about which we have no knowledge, and over which we have no control, it is natural to coarse-grain over these degrees of freedom. Such coarse-graining generally results in the recovery of approximate classical behaviour, and thus decoherence. We therefore anticipate that decoherence of histories corresponding to arrival times may be achieved for a generic state, provided that particle is coupled to a suitable environment. The second aim of this paper is to examine this scheme. We show that the probabilities calculated in this way are approximately decoherent, and approximately equal to those computed via the analogue of Eq.(\ref{1}), valid in the case of an environment.

Once we introduce an environment, however, the correspondence is  with a classical stochastic theory, rather than with a deterministic one. The classical trajectories may now cross the origin many times due to fluctuations. The arrival time distribution for such a classical theory was computed in Ref.\cite{HaZa}, and is given by
\beq
p(t_{1},t_{2})=\int_{t_{1}}^{t_{2}}dt\int_{-\infty}^{0} dp \int_{-\infty}^{\infty} dq \frac{(-p)}{m}\delta(q) w_{t}^{r}(p,q)\label{stoat},
\eeq
where $w_{t}^{r}(p,q)$ is the initial phase space distribution, evolved with a type of restricted propagator valid in the case of an environment, and with boundary conditions, \cite{HaZa}
\beq
w_{t}^{r}(p,0)=0,\quad {\rm for} \quad p>0 \label{wcond}.
\eeq

In Ref.\cite{HaZa} Halliwell and Zafiris presented a decoherent histories analysis of the corresponding quantum case. Although the conclusion they reach is sensible their analysis actually contains a small error. This was the result of a lack of appreciation of the role of the quantum Zeno effect in these calculations. We will show in this paper how the analysis can be corrected to give a correct expression for the arrival time probability in decoherent histories.  We shall be interested in a particular limit of this general case where the stochastic trajectories are sharply peaked about the deterministic trajectories we would have in the absence of any environment. In this case the restricted propagator may be replaced with an unrestricted one, and we will show that quantisation in this limit yields an expression of the form Eq.(\ref{1w}), but with the Wigner function evolved in the presence of an environment. 

It will eventually be necessary to specialise to a particular model of system-environment coupling in order to obtain quantitative results, but we will begin by considering general models which can be written in terms of a master equation of Lindblad form.  This kind of system has been extensively studied in the decoherence literature, and the properties of such evolution, such as suppression of interference effects, loss of entanglement and an approximate recovery of classical behaviour have been discussed in \cite{DoHa, HaZo, HaZo1, Zu}. 

This paper is arranged as follows, in Section \ref{sec2} we explore some properties of the arrival time distribution for general open systems of Lindblad form, our aim being to derive the corrections to the current resulting from the environmental dynamics. In Section \ref{sec3} we discuss quantum Brownian motion and specialise the results of Section \ref{sec2} to this case. In Section \ref{sec5} we derive some properties of the arrival time distribution for quantum Brownian motion, and in particular we prove that current becomes positive after a finite time. In Sections \ref{sec6} and \ref{sec7} we discuss the decoherent histories approach to quantum theory, and the expressions for the arrival time distribution that arise in this context. In Section \ref{sec8} we then examine whether histories corresponding to arriving in different intervals of time are decoherent. We summarise our results in Section \ref{sec9}.

\section{Arrival time for open quantum systems}\label{sec2}

We consider an open quantum system consisting of a free particle coupled to an environment with master equation of the following Lindblad form \cite{lind}, 
\beq
\frac{\partial \rho}{\partial t}=-\frac{i}{\hbar}[H,\rho]-\frac{1}{2}\sum_{j}\left(\{L_{j}^{\dagger}L_{j},\rho\}-2L_{j}\rho L_{j}^{\dagger}\right)\label{master},
\eeq
where $L_{j}$ are the Lindblad operators, and $H=H_{0}+H_{1}$ is the free hamiltonian plus a possible extra term that depends on the $L_{j}$. (This term gives rise to frequency renormalisation in oscillating systems.) Specific forms for the $L_{j}$ may be computed for particular models, but for the moment we leave this general. The Lindblad equation represents the most general master equation for a Markovian system which preserves the properties of the density matrix, in particular positivity.

We now extend the derivation of Eq.(\ref{1}) to this system. The probability of crossing during the interval $[t_{1},t_{2}]$ is
\bea
p(t_{1},t_{2})&=&\Tr( P \rho_{t_{1}})-\Tr( P \rho_{t_{2}})=\int_{t_{1}}^{t_{2}}dt \Tr( P\dot \rho_{t})\nonumber \\
&=&\frac{-1}{2m} \int_{t_{1}}^{t_{2}}dt \Tr \left([\hat p \delta(\hat x)+\delta(\hat x)\hat p]\rho_{t} \right)-\frac{i}{\hbar}\int_{t_{1}}^{t_{2}}dt \Tr \left( [H_{1}, P]\rho_{t}\right)\nonumber \\
&& +\frac{1}{2}\sum_{j}\int_{t_{1}}^{t_{2}}dt\Tr\left([L_{j}^{\dagger}, P] L_{j}\rho_{t}+L_{j}^{\dagger}[ P, L_{j}]\rho_{t} \right)\label{curo},
\eea
where $ P=\theta(\hat x)$. The first term is the standard current expression, although with the state evolved according to Eq.(\ref{master}), and we therefore recover Eq.(\ref{1}) when all the $L_{j}$ are 0. The second and third terms depend on the Lindblad operators, $L_{j}$ and thus on the form of the system-environment coupling.

To proceed further we specialise to the case where $L$ is a linear combination of $\hat x$ and $\hat p$, $L=a\hat x+i b \hat p$, where $a$ and $b$ are real constants. 
The master equation Eq.(\ref{master}) is then
\beq
\frac{\partial \rho_{t}}{\partial t}=-\frac{i}{\hbar}[H_{0},\rho_{t}]-iab[\hat x,\{\rho_{t},\hat p\}]-\frac{a^{2}}{2}[\hat x,[\hat x,\rho_{t}]]-\frac{b^{2}}{2}[\hat p,[\hat p,\rho_{t}]]\label{me}.
\eeq
Note that this equation is also identical in form to the exact master equation for a particle in a gas environment, given in \cite{Dme}.

To derive the arrival time distribution we could simply substitute $L=a\hat x+ib\hat p$ into Eq.(\ref{curo}), but the algebra is somewhat clumsy, and there are a number of terms which vanish for reasons not immediately apparent from this expression. An equivalent approach is to start from Eq.(\ref{me}), multiplying this expression by $ P$ and taking the trace. Because the second and third terms on the right of this expression have the form $[\hat x,\hat A]$ their contribution is proportional to
\beq
\Tr([\hat x,\hat A] P)=\Tr(\hat A[ P,\hat x])=0,\nonumber,
\eeq
so that only the first and last terms on the right hand side of Eq.(\ref{me}) contribute. The final term on the right of Eq.(\ref{me}) gives a contribution,
\beq
-\frac{b^{2}}{2}\Tr([\hat p,[\hat p,\rho_{t}] P)=-\frac{b^{2}}{2}\Tr([\hat p,[ P,\hat p]]\rho_{t})=-i\hbar b^{2}\Tr([\hat p \delta(\hat x)-\delta(\hat x)\hat p]\rho_{t}),
\eeq
so we arrive at the expression,
\bea
p(t_{1},t_{2})&=& \frac{-1}{2m}\int_{t_{1}}^{t_{2}}dt \Tr \left([\hat p \delta(\hat x)+\delta(\hat x)\hat p]\rho_{t} \right)\nonumber \\
&&-i\hbar b^{2}\int_{t_{1}}^{t_{2}}dt \Tr \left([\hat p \delta(\hat x)-\delta(\hat x)\hat p]\rho_{t} \right)\label{curL}.
\eea

The second term in this expression has a somewhat unusual form and its significance is not immediately clear. We will see below that this term is related to diffusion in position. There is a connection here to a recent paper by Genkin, Ferro and Lindroth \cite{Genkin}, in which the authors sought to examine the effects of an environment on the arrival time distribution. In that paper the authors used the standard expression for the current valid in the unitary case, Eq.(\ref{1}), ignoring the extra terms that arise because of the environment. Although this may be a good approximation when we can neglect the effects of dissipation, it is clear that there may be significant corrections to the current for strong dissipation. They also pointed out that these corrections are equivalent to the presence of extra terms in the continuity equation, although they did not compute these explicitly. For the sake of completeness, and also because it helps to understand the nature of the extra terms in Eq.(\ref{curL}) we will derive them here.

To derive the continuity equation we multiply Eq.(\ref{me}) by $\delta(\hat x-x)$ and perform the trace. If we neglect the final three terms on the right we arrive at the standard result,
\beq
\frac{\partial \rho_{t}}{\partial t}(x,x)+\frac{\partial J}{\partial x}(x,t)=0,
\eeq
with $J(x,t)$ defined by an obvious extension of Eq.(\ref{2}). Turning to the extra terms that result from the inclusion of the environment, the second and third terms vanish in exactly the same way as for the current above, and the correction term is therefore given by,
\beq
-\frac{b^{2}}{2}\Tr([\hat p,[\hat p,\rho_{t}]]\delta(\hat x-x))=2\hbar^{2}b^{2}\frac{\partial^{2} \rho_{t}}{\partial x^{2}}(x,x),
\eeq
so the continuity equation now reads,
\bea
\frac{\partial \rho_{t}}{\partial t}(x,x)+\frac{\partial}{\partial x}\left(J(x,t)+J_{D}(x,t)\right)&=&0,
\eea
where we have identified the diffusive current,
\beq
J_{D}(x,t)=-2\hbar^{2}b^{2}\frac{\partial \rho_{t}}{\partial x}(x,x).
\eeq 
This is a specific example of a modification to a conservation equation resulting from open system dynamics. For a more general discussion of these issues see \cite{Cos}.

The analysis presented above can also be phrased in terms of phase space distributions. The Wigner function corresponding to $\rho_{t}$ is defined as \cite{Wig}
\beq
W_{t}(p,q)=\frac{1}{2\pi\hbar}\int d\z e^{-\frac{i}{\hbar} p \z}\rho_{t}(q+\z/2,q-\z/2).
\eeq 
The Wigner transform of the master equation Eq.(\ref{me}) is
\beq
\frac{\partial W_{t}}{\partial t}=-\frac{p}{m}\dif{W_{t}}{q}+2\hbar^{2}ab\dif{(p W_{t})}{p}+\frac{\hbar^{2}a^{2}}{2}\diff{W_{t}}{p}+\frac{\hbar^{2}b^{2}}{2}\diff{W_{t}}{q}\label{mew}.
\eeq
The first term is the standard unitary term, whilst the second term represents dissipation, and the third and fourth terms represent  diffusion. 
The arrival time distribution can be written in terms of the Wigner function by taking the Wigner transform of Eq.(\ref{curL})
\beq
p(t_{1},t_{2})=\int_{t_{1}}^{t_{2}}dt\int dpdq\left(\frac{(-p)}{m}\delta(q)W_{t}(p,q)+\frac{\hbar^{2}b^{2}}{2}\delta(q) \dif{W_{t}}{q}(p,q)\right)\label{probwe}.
\eeq
Eqs.(\ref{curL}) and (\ref{probwe}) are the sought for generalisation of Eq.(\ref{1w}) for the case of a particle coupled to an environment.

\section{quantum Brownian motion}\label{sec3}
\label{sec4}
Quantum Brownian motion \cite{CaLe, HaZo, HaZo1} is a commonly used form for the environment of an open system, partly because it is exactly solvable, and partly because in many cases it is a good approximation to a realistic environment. For the quantum Brownian motion model \cite{HaZo} we have Eq.(\ref{me}) with 
\beq
a=\sqrt{2D/\hbar^{2}},\quad b=\frac{\gamma}{\sqrt{2D}}.
\eeq
 Here $D=2m\gamma kT$, where $T$ is the temperature of the environment, and $\gamma$ is a phenomenological damping constant, see \cite{CaLe}. 
Eq.(\ref{me}) may be written as
\bea
\frac{\d \rho_{t}}{\d t}(x,y)&=&\frac{i\hbar}{2m}\left(\frac{\d^{2}}{\d x^{2}}-\frac{\d^{2}}{\d y^{2}}\right)\rho_{t}(x,y)-\frac{D}{\hbar^{2}}(x-y)^{2}\rho_{t}(x,y)\nonumber\\
&&-\gamma(x-y)\left(\dif{}{x}-\dif{}{y}\right)\rho_{t}(x,y)-\frac{\hbar^{2}\gamma^{2}}{D}\left(\dif{}{x}-\dif{}{y}\right)^{2}\rho_{t}(x,y).
\eea

Although we could in principle work with this general case, it is useful to specialise to the case of negligible dissipation. There are two reasons for this, the first is that the analysis is considerably simplified, and this helps us to see the important effects more clearly. The second is that we have a particular aim in mind here, and that is to understand how a sensible classical result emerges from the quantum case. The classical case we have in mind is that of a heavy particle following an essentially classical, deterministic trajectory, but subject to small quantum fluctuations. We therefore restrict our analysis to timescales much shorter than the relaxation time $\gamma^{-1}$, and so we can drop the final two terms in the master equation above. 

This master equation may then be solved in terms of the propagator \cite{HaZo1},
\bea
\rho_{t}(x,y)&=&\int dx_{0}dy_{0}J(x,y,t|x_{0},y_{0},0)\rho_{0}(x_{0},y_{0})\\
J(x,y,t|x_{0},y_{0},0)&=&\left(\frac{m}{2\pi\hbar t}\right)\exp\left(\frac{im}{2\hbar t}\left[(x-x_{0})^{2}-(y-y_{0})^{2}\right]\right.\nonumber\\
&&\left.-\frac{Dt}{3}\left[(x-y)^{2}+(x-y)(x_{0}-y_{0})+(x_{0}-y_{0})^{2}\right]\right)\label{qbprop}.
\eea
Taking the same limit in the equation for the Wigner function, Eq.(\ref{mew}), gives
\beq
\frac{\d W_{t}}{\d t}=-\frac{p}{m}\frac{\d W_{t}}{\d q}+D\frac{\d^{2}W_{t}}{\d p^{2}}.
\eeq
Evolution of the Wigner function may also expresed in terms of a propagator \cite{HaZo1},
\bea
W_{t}(p,q)&=&\int dp_{0} dq_{0} K(q,p,t|q_{0},p_{0},0)W_{0}(p_{0},q_{0})\label{wprop}\\
K(q,p,t|q_{0},p_{0},0)&=&N \exp\left( -\a(p-p_{0})^{2}-\b \left(q-q_{0}-\frac{p_{0}t}{m}\right)^{2}\right.\nonumber \\ 
&&\left.+\e(p-p_{0})\left(q-q_{0}-\frac{p_{0}t}{m}\right)\right)\label{prop},
\eea
where $N,\: \a, \: \b,$ and $\e$ are given by
\beq
\a=\frac{1}{D t},\: \b=\frac{3m^{2}}{D t^{3}}, \: \e=\frac{3m}{D t^{2}}, \: N=\left(\frac{3m^{2}}{4 \pi^{2
} D^{2}t^{4}}\right)^{1/2}\label{prop2}.
\eeq

For later convenience we note that we can make the simple change of variables here, $q_{0}\to q_{0}-p_{0}t/m$, so that
\bea
W_{t}(p,q)&=&\int dp_{0} dq_{0} \tilde K(q,p,t|q_{0},p_{0},0)\tilde W_{0}(p_{0},q_{0})\label{wprop2}\\
\tilde K(q,p,t|q_{0},p_{0},0)&=&N \exp\left( -\a(p-p_{0})^{2}-\b \left(q-q_{0}\right)^{2}+\e(p-p_{0})(q-q_{0})\right)\nonumber\\
\tilde W(p,q)&=&W(p,q-pt/m)\label{wrdef}.
\eea
This form makes it clear that the evolution consists of two effects. The first is a shifting along the classical trajectories, whilst the second is a spreading in phase space. 

It is useful to consider this process in more detail. In the presence of an environment the width of the momentum distribution becomes time dependent, and we have,
\beq
(\Delta p)_{t}^{2}=(\Delta p)_{0}^{2}+Dt\nonumber, 
\eeq 
where $(\Delta p)_{0}$ is the momentum width of the initial state. We recognise an immediate difficulty here. Even for initial distributions consisting entirely of left moving momenta, $W_{t}$ will develop support on $p>0$ under evolution. This means we cannot strictly regard the current as an arrival time distribution, since a typical trajectory will now cross the origin many times. Differently put, the arrival time distribution is strictly a measurement of the first passage time, and this is no longer equal to the current. (Note that although this result is similar to the problems created by backflow, the reasons behind it are very different. The spreading of momentum induced by evolution in an environment is a purely classical effect.)

However, all is not lost. Although the current is no longer strictly the arrival time distribution, it may be a very good approximation to it. This is because the deviation of the current from the ``true'' arrival time distribution will be related to the probability that the state has the ``wrong'' sign momenta. This means that, provided we are in the ``near-deterministic'' limit,
\beq
(\Delta p)_{t}^{2}<<p_{0}^{2},
\eeq 
where $p_{0}$ is the momentum around which the initial state is tightly peaked, the current will still be a very good approximation to the true arrival time distribution. The timescale on which this analysis breaks down is given by the ``stochastic'' time,
\beq
\t_{s}=p_{0}^{2}/D\nonumber.
\eeq
After this time we must therefore revert to using the exact expression for the arrival time given by Eq.(\ref{stoat}). We see from the definition of $D$ that,
\beq
\t_{s}\gamma =\frac{p_{0}^{2}}{2m kT}>>1
\eeq
for the states we are interested in. This means the stochastic time $\t_{s}$ is much longer than the relaxation time $\gamma^{-1}$, so that working in the near-deterministic limit does not impose any further constraints compared with neglecting dissipation.

Returning to our arrival probability, in the limit of negligible dissipation Eq.(\ref{curL}), becomes,
\bea
p(t_{2},t_{1})&=&\int_{t_{1}}^{t_{2}}dt J(t)\label{cat}\\
J(t)&=&\int dq dp \frac{(-p)}{m}\delta(q) W_{t}(p,q)\label{3.3}.
\eea
This expression is now identical to the unitary case, Eq.(\ref{1}), but with the Wigner function evolved under quantum Brownian motion. 

We now turn to the question of whether inclusion of an environment helps ensure the positivity of Eq.(\ref{cat}).

\section{Properties of the arrival time distribution in quantum Brownian motion}\label{sec5}

In the introduction we noted that evolution in an environment typically renders $W_{t}$ positive after a short time. We now wish to examine the effect of this on our candidate arrival time distribution Eq.(\ref{cat}). To this end we introduce the notation \cite{DoHa}
\beq
\Z=\begin{pmatrix}p\\q \end{pmatrix}=\begin{pmatrix}z_{0}\\z_{1} \end{pmatrix},
\eeq
and also the class of Gaussian phase space functions
\beq
g(\Z;A)=\frac{1}{2\pi|A|^{1/2}}\exp\left(-\frac{1}{2}\Z^{T}A^{-1}\Z \right),
\eeq
where $A$ is a $2\times 2$ positive definite matrix, with determinant $|A|$. $g(\Z;A)$ will be a Wigner function if and only if 
\beq
|A|\geq \frac{\hbar^{2}}{4}.
\eeq
A useful result is that
\beq
\int d^{2}\Z g(\Z_{1}-\Z;A)g(\Z-\Z_{2};B)=g(\Z_{1}-\Z_{2}; A+B)\label{4.1}.
\eeq
In this notation we can write the propagation of the Wigner function, Eq.(\ref{wprop}), as 
\beq
W_{t}(\Z)=\int d\Z' g(\Z-\Z';A)\tilde W_{0}(\Z')\label{wpropg},
\eeq
where
 \beq
A=Dt   \begin{pmatrix} 
      2 & t/m \\
      t/m & 2t^{2}/3m^{2} \\
   \end{pmatrix}.
\eeq
Since 
\beq
|A|=\frac{D^{2}t^{4}}{3 m^{2}}\label{deta}
\eeq
after a time
\beq
t=\left(\frac{3}{16}\right)^{1/4}\left(\frac{2m\hbar}{D}\right)^{1/2}=\left(\frac{3}{16}\right)^{1/4}\t_{l}\label{ts}
\eeq
$g(\Z-\Z';A)$ will be a Wigner function, and thus $W_{t}(\Z)$ will be positive because it is equal to the convolution of two Wigner functions. Here $\t_{l}=\sqrt{2m\hbar/D}$ is the localisation time.

This is a useful result. Expressing the current Eq.(\ref{3.3}) in this new notation,
\beq
J(t)=\int d\Z \frac{(-z_{0})}{m}\delta(z_{1})W_{t}(\Z)\label{curj},
\eeq
we see that after the time given in Eq.(\ref{ts}), because $W_{t}>0$ the current Eq.(\ref{curj}) will be positive if the state consists purely of negative momenta. This means that after this time Eq.(\ref{cat}) will be a positive arrival time distribution. This holds provided times involved are much smaller than $\t_{s}$, as per the discussion below Eq.(\ref{wrdef}).

Now we turn to examining the properties of the current. We wish to find an expression for the current in the form $\Tr(P \rho)$, where $P$ is a projector or POVM. This would allow us to express the heuristic  arrival time distribution, Eq.(\ref{1w}), in the same form as other probabilities in quantum theory. Starting from Eq.(\ref{curj}) it is useful to write,
\bea
J(t)&=&\int d\Z\int d\Z' \frac{(-z_{0})}{m}\delta(z_{1}) g(\Z-\Z';A)\tilde W_{0}(\Z'),
\eea
using Eq.(\ref{wpropg}). 
We can deconvolve the propagator into two gaussians using Eq.(\ref{4.1}), in particular we will let $A=A_{0}+B$, where $A_{0}$ is a minimum uncertainty gaussian, and $B$ is the remainder.
\beq
A_{0}= \hbar  \begin{pmatrix} 
      s^{2} & 0 \\
      0& 1/4s^{2} \\
   \end{pmatrix}
\eeq
\beq
B=   Dt   \begin{pmatrix} 
      2 & t/m \\
      t/m & 2t^{2}/3m^{2} \\
   \end{pmatrix} - A_{0}.
\eeq
Here $s$ is some real number.
Using the convolution property Eq.(\ref{4.1}) we can write the current Eq.(\ref{curj}) as
\bea
J(t)&=& \int d\Z'' \left[\int d \Z \frac{(-z^{0})}{m}\delta(z^{1})g(\Z-\Z'';B)\right]\left[\int d\Z'\tilde W_{0}(\Z')g(\Z''-\Z';A_{0})\right]\nonumber\\
&=& \int d\Z'' \left[\int d \Z \frac{(-z^{0})}{m}\delta(z^{1}+z_{0}t/m)g(\Z-\Z'';B)\right]Q(\Z'')\label{4a},
\eea
where $Q(\Z)$ is the Q-function \cite{Qfn}, and we have undone the change of variables implied in Eq.(\ref{wrdef}). The Q-function can be written as
\beq
Q(\Z)=\frac{1}{\pi}\bra{\Z}\rho \ket{\Z}\nonumber.
\eeq
We can therefore express the current as,
\beq
J(t)=\Tr(F \rho)\label{opcur},
\eeq
where
\bea
F&=&\frac{1}{\pi} \int d\Z'' \ket{\Z''}\bra{\Z''}\left(\int d\Z \frac{(-z_{0})}{m}\delta(z_{1}+z_{0}t/m)g(\Z-\Z'';B)\right)\nonumber\\
&=&\int d\Z P_{\Z}\frac{(-z_{0})}{m}\delta(z_{1}+z_{0}t/m),
\eea
and we have defined the POVM
\beq
P_{\Z}=\frac{1}{\pi}\int d\Z'' \ket{\Z''}\bra{\Z''}g(\Z-\Z'';B),
\eeq
which is clearly a phase space operator localised around $\Z$. $F$ is therefore a smeared version of the  object used to compute the current classically, $-p\delta(x_{t})/m$.
This holds for times
\beq
\left(\frac{3}{16}\right)^{1/4}\t_{l}\leq t<<\t_{s}.
\eeq

Assume for a moment that $B=0$, and so $P_{\Z}=\ket{\Z}\bra{\Z}$. The current would then be given by
\beq
J(t)=\int d\Z \frac{(-z_{0})}{m}\delta(z_{1}-z_{0}t/m)Q(\Z)\label{crq}.
\eeq
Since the Q-function is positive by construction the current computed in this way will be positive to the extent that the Q-function has vanishing support on $p>0$ ($Q(\Z)$ cannot strictly vanish for $p>0$ but it can be exponentially small, and this suffices here. See the comments below Eq.(\ref{wrdef}).) For $B>0$ the Q-function is simply smeared further, and this will preserve the property of positivity. In fact one could imagine {\it postulating} Eq.(\ref{crq}) as the definition of the arrival time even in the unitary case, since it clearly satisfies all the conditions required.

The probability of arriving in the interval $[t_{1},t_{2}]$ is given by the integral of Eq.(\ref{opcur}), and because $P_{\Z}$ is time dependent (via $B$) this will not in general have a simple form. However there is a separation of time scales here, the time scale of evolution of $P_{\Z}$ is seen from Eq.(\ref{deta}) to be $\t_{l}$. So, if $t_{2}-t_{1}<<\t_{l}$ we have approximately
\beq
p(t_{1},t_{2})=\int_{t_{1}}^{t_{2}} dt \Tr(F \rho)\approx \int d\Z\Tr\left[\rho P_{\Z} \int_{t_{1}}^{t_{2}}dt \frac{(-z_{0})}{m}\delta(z_{1}+z_{0}t/m)\right]=\Tr(E\rho)\label{atpovm},
\eeq
where 
\beq
E=\int d\Z P_{\Z}[\theta(z_{1}+z_{0}t_{1}/m)-\theta(z_{1}+z_{0}t_{2}/m)]\label{defE}
\eeq
is a POVM representing arrival at $x=0$ between $t_{1}$ and $t_{2}$.
Note the close similarity between this object and the operator Eq.(\ref{2}), but that crucially this phase space operator is positive for $z_{0}<0$ (ie $p<0$).
In this case therefore, the effect of the environment is simply to smear the unitary result over a region of phase space, given by $P_{\Z}$ computed at $t_{1}$. 

Eqs.(\ref{atpovm}) and (\ref{defE}) form the first significant result of this paper. Eq.(\ref{atpovm}) expresses the heuristic arrival time distribution, Eq.(\ref{1w}), as the trace of an operator times a POVM, and thus has the same form as standard expressions for probability in quantum theory. The POVM, Eq.(\ref{defE}), arises because the environment effectively measures the system.

We have therefore discovered a range of times for which the current, Eq.(\ref{curj}), gives a positive arrival time distribution. After a time of order $\t_{l}$ interference effects vanish and we can regard Eq.(\ref{curj}) as the arrival time distribution, which can also be written in the form Eq.(\ref{opcur}). Eventually, however, on a time scale of order $\t_{s}$ the environment causes diffusion in momentum to such an extent that the trajectories of the particle are no longer sharply peaked around the classical trajectory computed in the absense of an environment. Since the particle is still behaving classically after this time there will exist an arrival time distribution of the form Eq.(\ref{stoat}), but our simpler expression Eq.(\ref{curj}) will no longer be a good approximation to this. It is easy to show that $\t_{s}/\t_{l}=E \t_{l}\hbar$ where $E$ is the energy of the initial state, and thus these expressions are valid for an large interval if $E \t_{l}>>\hbar$.

\section{The decoherent histories approach to the arrival time problem} \label{sec6}

\subsection{The Decoherent Histories approach to Quantum Theory}
We begin by briefly reviewing the decoherent histories approach to quantum theory.  More extensive discussions can be found in the references \cite{GeH1, Gri, Omn, Hal2, DoH}. 

Alternatives at fixed moments of time in quantum theory are
represented by a set of projection operators $\{ P_a \}$,
satisfying the conditions
\bea
\sum_a P_a &=& 1\\
P_a P_b &=& \delta_{ab} P_a,
\eea 
where we take $a$ to run over some finite range. In the decoherent histories approach to quantum
theory histories are represented by class operators $C_{\a}$ which are time-ordered
strings of projections
\beq
C_{\a} = P_{a_n} (t_n) \cdots P_{a_1}(t_1)
\label{1.3}
\eeq
or sums of such strings  \cite{Ish}. Here the projections are in the Heisenberg
picture and $ \a $ denotes the string $ (a_1, \cdots a_n)$. 
All class operators satisfy the condition
\beq
\sum_{\a} C_{\a} = 1.
\label{1.4}
\eeq
Probabilities are assigned to histories via the formula
\beq
p(\a) = {\rm Tr} \left( C_{\a} \rho C_{\a}^{\dag} \right).
\label{1.6}
\eeq
However probabilities assigned in this way do not necessarily obey the
probability sum rules, because of quantum interference.
We therefore introduce the decoherence functional
\beq
D(\a, \b) = {\rm Tr} \left( C_{\a} \rho C_{\b}^{\dag} \right),
\eeq
which may be thought of as a measure of interference between pairs of histories.
We require that sets of histories satisfy the condition of
decoherence, which is
\beq
D(\a, \b) = 0, \ \ \ \a \ne \b\label{1.7}.
\eeq
This ensures that all probability sum rules are satisfied.

We note briefly that when there is decoherence Eq.(\ref{1.4}) and Eq.(\ref{1.7}) together imply that the probabilities $p(\a)$ are given by the simpler expressions
\beq
q(\a) = {\rm Tr} \left( C_{\a} \rho \right).
\label{1.17}
\eeq
Decoherence ensures that $q(\a)$ is real and positive, even though it
is not in general. 
In this way decoherent histories may reproduce probabilities of the form Eq.(\ref{2}).

\subsection{The Decoherent Histories aproach to the arrival time problem}

We turn now to the definition of the arrival time problem in decoherent histories. In Refs.\cite{HaYe1, HaYe2}  the class operators corresponding to a first crossing of the origin in the time interval $[t_{k-1}, t_{k}]$ were computed to be
\beq
C_k =  \bar P ( t_{k} ) P (t_{k-1}) \cdots P(t_2) P(t_1),
\label{24}
\eeq 
for $ k \ge 2 $, with $C_1 = \bar P (t_1)$ and where $\bar P(t)=\theta(-\hat x_{t})$. These clearly describe histories
which are in $ x>0$ at times $ t_1, t_2, \cdots t_{k-1}$ and in $x<0$ at
time $t_k$, so, approximately,
describe a first crossing between $t_{k-1}$ and $t_k$. We are prevented from taking the time between projections to zero by the Zeno effect \cite{Zeno}. This point was extensively discussed in \cite{HaYe1,HaYe2}.

In Ref \cite{HaYe2} these class operators were then simplified with the help of the following semiclassical approximation
\beq
 P (t_n) \cdots P(t_2) P(t_1) | \psi \rangle \approx P (t_n) | \psi \rangle.
\label{21}
\eeq
This is the statement that given that the state is in $x>0$ at $t_{n}$, it must also have been $x>0$ at all previous times (we restrict attention here to states with $p<0$). It is clear from the path integral representation of the propagator that this is only true semiclassically. From this we obtain the class operators for first crossing between $t_{k-1}$ and $t_k$ as
\bea
C_{k}& \approx & \bar P ( t_{k} ) P (t_{k-1})\label{ap1} \\
&=& P ( t_{k-1} ) - P ( t_{k})  P ( t_{k-1} )
\nonumber \\
& \approx & P ( t_{k-1} ) - P ( t_{k} )\label{ap2},
\eea
where we use the semiclassical approximation again to arrive at the last line. In the case of a free particle without an environment, this class operator reproduces Eq.(\ref{1}) under the conditions of decoherence. This is because, assuming decoherence, the arrival time probability computed from decoherent histories is
\beq
p(t_{2},t_{1})=\Tr(C\rho C^{\dagger})=\Tr(C \rho),
\eeq
using Eq.(\ref{1.17}).

\section{Decoherence and the semi-classical approximation for arrival time}\label{sec7}

In the previous section we introduced the arrival time problem in decoherent histories. The key step in making contact with the simple heuristic result of Eq.(\ref{1}) was the approximation of the class operators  Eq.(\ref{24}) by Eq.(\ref{ap2}). We now want to explore this approximation, and prove that it is indeed valid in some interesting limit.

For a free particle, the error in this approximation comes from  the finite width of the wave packet. This means that
\beq
P(t_{k})\ket{\psi}\neq \ket{\psi},
\eeq
 even for times, $t_{k}$, significantly before the classical crossing time. However it is clear that the largest error will come from the last projection that preceeds the classical crossing time. Since the spacing between projections has to be greater than the Zeno time, which is also the time scale on which a wavepacket may be said to ``cross'' the origin, only one or two projections stand any chance of significantly disturbing the wavepacket. 

Writing this in terms of the density matrix,
\bea
\Tr( P(t_{k+1})P(t_{k})\rho P(t_{k}))&=&\Tr(P(t_{k+1})\rho)-\Tr(P(t_{k+1})\bar P(t_{k})\rho)-\Tr(\rho \bar P(t_{k}) P(t_{k+1}))\nonumber \\
& &+\Tr(P(t_{k+1})\bar P(t_{k})\rho \bar P(t_{k}))\nonumber  \\
&\approx& \Tr(P(t_{k+1})\rho)\nonumber,
\eea
where the last line defines the semi-classical approximation. Noting that, with $t_{m}\geq t_{k}$
\beq
\left| \Tr(P(t_{m}) \bar P(t_{k})\rho)\right|^{2}\leq \Tr(P(t_{m}) \bar P(t_{k})\rho \bar P(t_{k})).
\eeq A sufficient condition for the validity of the semi-clasical approximation is therefore that the object
\beq
\Delta_{k,m} := \Tr(P(t_{m})\bar P(t_{k})\rho \bar P(t_{k}))\label{delta}
\eeq
is much smaller than 1 for all $t_{m}>t_{k}$.

Now assume for a moment that the semiclassical approximation holds, so that the class operators are given by Eq.(\ref{ap2}). The  probability of first crossing between $t_{k}$ and $t_{k+1}$ is given by
\bea
\Tr(C_{k}\rho C^{\dagger}_{k})&=& \Tr(C_{k}\rho)+2\Tr(\rho  P(t_{k+1})))-\Tr(\rho P(t_{k}) P(t_{k+1})))-\Tr(P(t_{k+1})P(t_{k})\rho)\nonumber.
\eea
However since we are assuming that the semi-classical approximation holds, the last three terms cancel and we obtain
\beq
\Tr(C_{k}\rho C^{\dagger}_{k})=\Tr(C_{k}\rho)=\Tr(\rho C^{\dagger}_{k})\nonumber.
\eeq
Furthermore, consider a general off-diagonal term in the decoherence functional, where without loss of generality we take $t_{k+1}<t_{m}$
\bea
\Tr(C_{k}\rho C^{\dagger}_{m})&=&\Tr(  P(t_{m}) P(t_{k})\rho)-\Tr( P(t_{m}) P(t_{k+1})\rho)\nonumber\\
&&-\Tr( P(t_{m+1}) P(t_{k})\rho)+\Tr( P(t_{m+1}) P(t_{k+1})\rho).
\eea
We see that this vanishes if the semi-classical approximation holds.

What we have shown therefore, is that the semi-classical condition and the decoherence conditions are closely related. If the object given in Eq.(\ref{delta}) is much smaller than 1 then both conditions are satisfied. This means firstly that the class operators are approximately given by Eq.(\ref{ap2}), and secondly that the histories described by these class operators are approximately decoherent.

In general, the probabilities for time of arrival as computed by decoherent histories differ from the heuristic ones obtained by the standard quantum mechanical analysis. We don't expect agreement since, as discussed in the Introduction, the heuristic formula, Eq.(\ref{1}), is not in general of the canonical form required for genuine quantum mechanical probabilities. It is also the case that decoherent histories only ascribes probabilities to certain sets of histories. However if the object $\Delta_{k,m}$ is small, then we have shown that the decoherent histories analysis reproduces the probabilities computed in the standard, heuristic, way.

As well as clarifying the analysis of \cite{HaYe2}, note that all of the statements above apply to the case of a particle coupled to an environment. In this case the trace is to be taken over the system and environment, and $P(t_{k})=P(t_{k})_{s}\otimes 1_{\e}$ is a projection onto the degrees of freedom of the system, $s$, tensored with the identity operator on the environmental degrees of freedom, $\e$. This is the case of interest in this paper.

Since the object $\Delta_{k,m}$ plays a central role in our analysis it is interesting to ask if it has any physical interpretation. The answer is that it does. Recall that our class operator for crossing between $t_{k-1}$ and $t_{k}$ is given by Eq.(\ref{ap2}). We see however that
\beq
P(t_{k-1})-P(t_{k})=\bar P(t_{k})P(t_{k-1})-P(t_{k})\bar P(t_{k-1})\label{kterms}.
\eeq
The approximation leading from Eq.(\ref{ap1}) to Eq.(\ref{ap2}) is equivalent to dropping the second term on the right hand side of Eq.(\ref{kterms}). The error in this approximation can be estimated by computing the probability associated with the class operator  $P(t_{k})\bar P(t_{k-1})$. We see
\beq
\Tr([P(t_{k})\bar P(t_{k-1})]\rho[P(t_{k})\bar P(t_{k-1})]^{\dagger})=\Tr(P(t_{k})\bar P(t_{k-1})\rho \bar P(t_{k-1}))=\Delta_{k-1,k}\nonumber.
\eeq
However Eq.(\ref{kterms}) has a simple physical interpretation: it is the decomposition of the total current into right and left moving parts. The object $\Delta_{k,m}$ is therefore just the probability associated with the right moving current. Classically this is small by construction, since we will choose to work with wavepackets tightly peaked in negative momentum. Quantum mechanically however this term need not be small, indeed the existence of the backflow effect \cite{BrMe} shows that this term can sometimes be larger than the left moving current, even for wavepackets composed entirely of negative momenta. 

The semi-classical condition we are imposing, and by extension the decoherence condition, is therefore stronger than the condition for the current to be positive. Decoherence requires the absence of interference between crossings at different times, whilst the standard heuristic analysis includes some interference effects, provided they are not so large as to render the arrival time probabilities negative. The significance of this will be explored elsewhere.

\section{Decoherence of arrival times in quantum Brownian motion}\label{sec8}

\subsection{General case}

In the previous section we have shown that an arrival time distribution may be derived from decoherent histories, and that it agrees with the current provided there is decoherence. We therefore turn now to the question of determining for which states and intervals decoherence is achieved. Recall that our aim is to show that the inclusion of an environment gives rise to decoherence of arrival time probabilities for generic initial states of the system.
We will start our decoherent histories analysis with a discussion of the general case considered in Ref.\cite{HaZa}, but most of our detailed results will concern the near-deterministic limit discussed in the introduction, and in the discussion of the Wigner function above.
 
The decoherence functional may be written in path integral form as
\beq
\D(\a,\a')=\int_{\a} \D x \int_{\a'} \D y \exp\left(\frac{i}{\hbar} S[x]- \frac{i}{\hbar}S[y]+iW[x,y]\right) \rho_{0}(x,y)\label{3.1},
\eeq
where $\a,\a'$ represent the restriction to paths that are, for example, in $x>0$ at $t_1$ and in $x<0$ at $t_{2}$.  $W[x,y]$ is the Feynman-Veron influence functional phase which summarises the effect of the environment, and is given in the case of negligible dissipation by
\beq
W[x,y]=\frac{i D}{\hbar^{2}}\int dt(x-y)^{2}.
\eeq 
It is this phase which is responsible for the suppression of interference between paths $x(t)$ and $y(t)$ that differ greatly, and produces decoherence. 

The decoherent histories analysis we are about to perform was first attempted by Halliwell and Zafiris in Ref.\cite{HaZa}. Their conclusions are reasonable, but the analysis leading to them in fact contains a small error. This error arose due to a lack of appreciation of the role of the Zeno effect, as discussed in the Introduction. We aim to show here how the analysis may be modified in line with the treatment of the arrival time problem presented here and in Ref.\cite{HaYe2}.

We start from the expression for the density matrix for states that do not cross the origin in a time interval $[0,t]$ (Ref.\cite{HaZa}, Eq.(4.34))
\beq
\rho_{rr}(x_{f},y_{f})=\int_{r}\D x\int_{r}\D y \exp\left[\frac{im}{2\hbar}\int dt (\dot x^{2}-\dot y^{2})-\frac{D}{\hbar^{2}}\int dt(x-y)^{2}\right]\rho_{0}(x_{0},y_{0}).
\eeq
Here the restriction is interpreted as $x,y>0$ at times $\e, 2\e...$. The probability that the state does not cross the origin in this interval is then given by the trace of this expression. There are two interesting limits to be taken here. The first is letting $\e\to0$, and this recovers the notion of the restricted propagator. The second limit is that of strong decoherence effects, so that we can assume that the path integral is tightly peaked in $(x-y)$, this recovers the classical limit. The claim in Ref.\cite{HaZa} is that we can take these limits simultaneously. To see why this is problematic we write the restricted path integral as a product of propagators,
\beq
\rho_{rr}(x_{f},y_{f})=\prod_{k=0}^{n}\int_{0}^{\infty}  dx_{k} dy_{k}J(x_{k+1},y_{k+1},t_{k+1}|x_{k},y_{k},t_{k})\rho_{0}(x_{0},y_{0}),
\eeq
where $t_{k}=k\e$, $t_{0}=0$, $t_{n+1}=t$.
Changing variables to $X=x+y$, $\z=\frac{1}{2}(x-y)$ gives
\bea
\rho_{rr}(x_{f},y_{f})&=&\prod_{k=0}^{n} \int_{0}^{\infty} dX_{k} \int_{-X_{k}}^{X_{k}} d\z_{k} \left(\frac{m}{\pi\hbar \e}\right)\exp\left[\frac{im}{\hbar \e}[(\z_{k+1}-\z_{k})(X_{k+1}-X_{k})]\right. \nonumber\\
&&\left.-\frac{D\e}{3\hbar^{2}}[\z_{k+1}^{2}+\z_{k+1}\z_{k}+\z_{k}^{2}]\right]\rho_{0}(X_{0},\z_{0}).
\eea
Now in order for the product of propagators to be equal to a restricted propagator, we need to take $\e\to0$. However to recover the classical result we need to replace the limits of integration for $\z_{k}$ with $\pm \infty$, which requires that $D\e>>\hbar^{2}/l^{2}$ where $l$ is some length scale. Clearly these two limits are incompatible for finite $D$.

The standard discussion of the Zeno effect emphasizes the role of the so called ``Zeno time'', $\t_{z}=\hbar(\Delta H)^{-1}$. It is generally held that projections separated by times greater than this will not give rise to the Zeno effect. The issue is not without subtlety in the present case however, firstly because the Zeno effect is usually discussed in the context of unitary evolution, and this is not the case here, and secondly because characteristic times associated with the Zeno effect usually refer to selective measurement, and a projection onto positive $x$ is certainly non-selective. This casts doubt on the role of the Zeno time as a timescale in this problem. Studies of the arrival time problem defined using a complex potential \cite{HaYe1} suggest that the relevant timescale might be given by $\hbar/E$. This issue will be taken up elsewhere \cite{HaYe3}.

In any case it is clear that there exists a short timescale on which we can still approximate the limits in the $\z$ integrals above in the manner indicted, but that is short enough to give non-trivial crossing probabilities. Following the steps in Ref.\cite{HaZa} then gives 
\beq
p(t_{1},t_{2})=1-\Tr(\rho_{rr})=\int_{t_{1}}^{t_{2}}dt\int_{-\infty}^{0} dp \int_{-\infty}^{\infty} dq \frac{(-p)}{m}\delta(q) W_{t}^{r}(p,q)\label{stoat2},
\eeq
where the Wigner function, $W^{r}_{t}$ obeys,
\beq
W_{t}^{r}(p,q)=\int dp_{0} dq_{0}K_{r}(p,q,t|p_{0},q_{0},0) W_{0}(p_{0},q_{0}),
\eeq
where $K_{r}$ is the ``restricted'' propagator, defined by
\beq
K_{r}(p,q,t|p_{0},q_{0},0)=\prod_{k=1}^{n}\int dp_{k}\int _{0}^{\infty}dq_{k}K(p_{k},q_{k},t_{k}|p_{k-1},q_{k-1},t_{k-1}),
\eeq
with $t_{k}=k\e$. However because we are now dealing with an essentially classical system, this propagator is well approximated by its continuum limit. 
Understood in this sense, we see that the conclusions of Ref.\cite{HaZa} are in fact correct, even if the argument leading to them is not. The interesting subtleties that arise in this argument relate to the classical limit of the Zeno effect, and this will be explored elsewhere \cite{HaYe3}.

As important as this general case is, it is of interest to examine the simpler case of near-deterministic evolution, where the classical limit of the arrival time probability is simply expected to be the time integral of the current density. The analysis in this case simplifies considerably, and we will explicitly exhibit decoherence of histories for suitable intervals.

\subsection{The near-deterministic limit}
We have previously seen that, in the near-deterministic limit, if there is decoherence then the arrival time probabilities derived from decoherent histories agree with those computed from the current. We therefore turn now to discussing the conditions under which we have decoherence of crossing probabilities.  The general picture we have in mind is illustrated in Fig.(1), we have an initial wavepacket defined at $t=0$, evolved in the presence of an environment until time $t_{1}$, and then we wish to compute the probability of crossing between $t_{1}$ and $t_{2}$.
\begin{figure}
    \centering
        \includegraphics[width=4.7in]{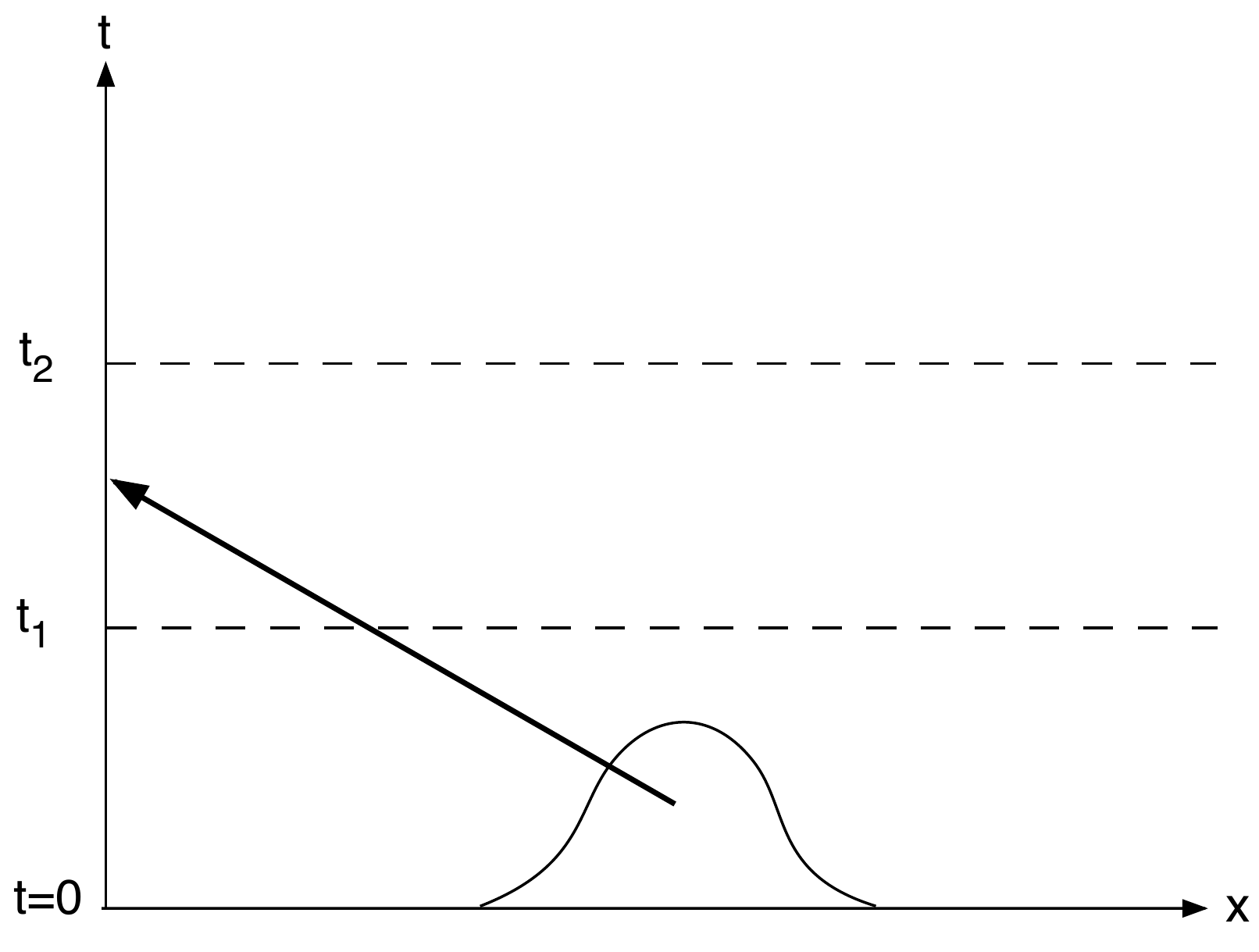}
    \caption{The probability that a generic wavepacket arrives between $t_{1}$ and $t_{2}$ can be expressed in terms of the state at $t_{1}$ or, via the propagator, in terms of an initial time $t=0$.}
    \label{fig:Quadratic}
\end{figure}

Our task is to compute the off-diagonal elements of Eq.(\ref{3.1}), and this will be rather involved in general. However we are saved from having to do this by the observation in Section \ref{sec7} about the relation between the semi-classical condition and the decoherence condition. It suffices therefore to compute the quantity $\Delta_{k,m}$ defined in Eq.(\ref{delta}), in the presence of an environment. We anticipate that this will be small simply from the form of Eq.(\ref{3.1}). This is because the effect of the environment is to cause the density matrix to become tightly peaked around the classical path and classically, for $p<0$ the probability given by $\Delta_{k,m}$ is zero. We will see how this works in a specific example below. 

We take $k=1, m=2$ without loss of generality, and we drop the subscript from now on
\beq
\Delta=\Tr(P(t_{2})\bar P(t_{1})\rho \bar P(t_{1}))=\int_{\a}\D x \int_{\a} \D y \exp\left(\frac{i}{\hbar} S[x]-\frac{i}{\hbar} S[y]- \frac{D}{\hbar^{2}}\int dt(x-y)^{2}\right) \rho_{t_{1}}(x_{1},y_{1}),
\eeq
where the histories $\a$ are those that start at $x_{1},y_{1}<0$ and finish at $x_{2}>0$. In terms of the density matrix propagator, Eq.(\ref{qbprop})
\bea
\Delta&=&\int_{0}^{\infty}dx_{2}\int_{-\infty}^{0}dx_{1}\int_{-\infty}^{0}dy_{1}J(x_{2},x_{2},t_{2}|,x_{1},y_{1},t_{1})\rho_{t_{1}}(x_{1},y_{1})\nonumber \\
&=& \int_{0}^{\infty}dx_{2}\int_{-\infty}^{0}dx_{1}\int_{-\infty}^{0}dy_{1} \left(\frac{m}{\pi\hbar t}\right)\nonumber \\ &&\exp\left(\frac{im}{2\hbar(t_{2}-t_{1})}((x_{2}-x_{1})^{2}-(x_{2}-y_{1})^{2})-\frac{D(t_{2}-t_{1})}{3\hbar^{2}}(x_{1}-y_{1})^{2}\right)\rho_{t_{1}}(x_{1},y_{1})\nonumber.
\eea
Transforming to new variables
\beq
X=\frac{x_{1}+y_{1}}{2},\quad \z=x_{1}-y_{1},
\eeq
and writing the density matrix in terms of the Wigner function via
\beq
\rho(x,y)=\int_{-\infty}^{\infty}dp e^{\frac{i}{\hbar}p(x-y)}W(p,\frac{x+y}{2}),
\eeq
we obtain
\bea
\Delta&=&-\int_{0}^{\infty}dx_{2}\int_{-\infty}^{0}dX\int_{-X}^{X}d\z \int_{-\infty}^{\infty}dp \left(\frac{m}{2\pi\hbar (t_{2}-t_{1})}\right) \nonumber \\ 
&&\exp\left(\frac{im}{\hbar(t_{2}-t_{1})}\z(X-x_{2})+\frac{i}{\hbar} p\z-\frac{D(t_{2}-t_{1})}{3\hbar^{2}}\z^{2}\right)W_{t_{1}}(X,p)\label{int}.
\eea
We see from this expression that there is a time scale, $\t_{l}=\sqrt{2m\hbar/D}$, set by the environment on which localisation effect are important. This timescale is the same as that on which the current becomes positive, as we saw earier.

There are three cases to explore here, the first is where there is no environment, $D=0$. This is the case covered in Ref.\cite{HaYe1}. 
The second case is the intermediate one, $t_{1}/\t_{l}>>1$ but $(t_{2}-t_{1})/\t_{l}<<1$. This is the most general case in which we can expect to have environmentally induced decoherence.
The final case is where $t_{1}/\t_{l}, (t_{2}-t_{1})/\t_{l}>>1$. This is the case of very strong environmental coupling.  

For the free case, $D=0$ the integrals over $\z$ and $x_{2}$ in Eq.(\ref{int}) may be carried out to give,
\beq
\Delta=\int_{-\infty}^{0}dX \int_{-\infty}^{\infty}dp W_{t_{1}}(X,p)f\left[\frac{X}{\hbar}\left(\frac{mX}{(t_{2}-t_{1})}+p\right)\right]\label{deltaw},
\eeq
where 
\beq
f(u)=\frac{1}{\pi}\int_{u}^{\infty}dy \frac{\sin{y}}{y}.
\eeq 
This expression for $\Delta$ is now identical to Eq.(5.43) of Ref.\cite{HaYe1}, and the same conclusions apply. We briefly repeat the analysis here however, since it turns out to be relevant for the second case. 

We note firstly that $f(+\infty)=0, f(-\infty)=1, f(0)=1/2$. Now we assume that our state at $t_{1}$ is of the form
\beq
W_{t_{1}}(X,p)=\frac{1}{\pi\hbar}\exp\left(-\frac{(X-X_{0}-p_{0}t_{1}/m)^{2}}{2\s^{2}}-\frac{2\s^{2}}{\hbar^{2}}(p-p_{0})^{2}\right),
\eeq
and that $\s^{2}$ is large, so the state is tightly peaked in momentum. We can therefore integrate out $p$, setting $p=p_{0}$ to obtain,
\beq
\Delta=\int_{-\infty}^{0}dX \frac{1}{\sqrt{2\pi\s^{2}}} \exp\left(-\frac{(X-X_{0}-p_{0}t_{1}/m)^{2}}{2\s^{2}}\right)f\left[\frac{X}{\hbar}\left(\frac{mX}{(t_{2}-t_{1})}+p_{0}\right)\right].
\eeq
In \cite{HaYe1} it was noted that this type of integral is dominated by values of $X$ for which
 \beq
 0\leq X\leq -\frac{\pi\hbar}{4|p_{0}|},
 \eeq
 providied that $p_{0}^{2}(t_{2}-t_{1})/2m>>1$, and in this range the exponential term is approximately constant. We can approximate this integral then by intergrating from $0$ to $-\pi/4|p_{0}|$, taking $X=0$ in the exponential term and approximating,
 \beq
 f(u)\approx \frac{1}{2}-\frac{u}{\pi}+O(u^{3}).
 \eeq
This gives
\beq
\Delta \approx \sqrt{\frac{\pi}{2}}\frac{\hbar}{8\s |p_{0}|}\exp\left(-\frac{(X_{0}+p_{0}t_{1}/m)^{2}}{2\s^{2}}\right)<<1,
\eeq
there will therefore be decoherence for gaussian wavepackets tightly peaked in momentum, provided their momentum is such that $p_{0}^{2}(t_{2}-t_{1})/2m>>\hbar$, ie $E_{0}(t_{2}-t_{1})>>\hbar$. In \cite{HaYe1} it was argued that this conclusion also holds for orthogonal superpositions of gaussians.

Turning to the intermediate case, since $(t_{2}-t_{1})/\t_{l}<<1$ we can set $D=0$ in Eq.(\ref{int}) and we obtain Eq.(\ref{deltaw}) again. Now however, $W_{t_{1}}(X,P)$ is the initial state evolved with an environment, and since $t_{1}/\t_{l}>>1$ this should be significant. To proceed we write the Wigner function at time $t_{1}$ in terms of the initial state at $t=0$ using Eq.(\ref{wprop})
\beq
\Delta= \int dX_{0} dp_{0} W_{0}(X_{0},p_{0}) F[X_{0},p_{0}]
\eeq
\beq
F[X_{0},p_{0}]=\int_{-\infty}^{0}dX  \int_{-\infty}^{\infty}dp K(X,p,t_{1}|X_{0},p_{0},0) f\left[\frac{X}{\hbar}\left(\frac{mX}{(t_{2}-t_{1})}+p\right)\right].
\eeq
Because the Wigner function propagator is a gaussian the analysis is similar to the first case. Recall that we are assuming $t<<\t_{s}$ and so $p_{0}^{2}/Dt_{1}>>1$, this means we can integrate out $p$, setting $p=p_{0}$. This gives us
\beq
F[X_{0},p_{0}]=\int_{-\infty}^{0}dX N\sqrt{\frac{\pi}{\a}} \exp\left(  -\frac{\b}{4}(X-X_{0}-p_{0}t_{1}/m)^{2})\right)f\left[\frac{X}{\hbar}\left(\frac{mX}{(t_{2}-t_{1})}+p_{0}\right)\right],
\eeq
where, $N,\a,\b$ are defined in Eq.(\ref{prop2}). Now,
\beq
p_{0}^{2}/Dt_{1}>>1
\eeq
implies that
\beq
\frac{1}{p_{0}^{2}}<<\frac{1}{\hbar^{2}\b}\left(\frac{\t_{l}}{t_{1}}\right)^{4},
\eeq
so that for $t_{1}>\t_{l}$, $1/p_{0}^{2}<<1/\b\hbar^{2}$. This means, again like the case of $D=0$, that the exponential term is roughly constant compared to $f$.  Since we have $p_{0}^{2}(t_{2}-t_{1})/2m>>\hbar$, we follow the same procedure as the $D=0$ case, arriving finally at
\beq
F[X_{0},P_{0}]\approx \frac{\sqrt{\pi}\hbar}{8|p_{0}|}\sqrt{\frac{\b}{4}}\exp\left(-\frac{\b}{4}(X_{0}+p_{0}t_{1}/m)^{2}\right).
\eeq
The value of $\Delta$ now depends on the relationship between the width of the initial state $\s$ and $\b$. However we can obtain an upper bound by ignoring the effects of the exponential term in $F$, this gives
\beq
\Delta \leq\frac{\sqrt{\pi}\hbar}{8|p_{0}|}\sqrt{\frac{\b}{4}}=\frac{1}{16}\sqrt{\frac{2m\hbar}{p_{0}^{2}t_{1}}}\left(\frac{\t_{l}}{t_{1}}\right)<<1.
\eeq

Finally we have the case of strong system-environment coupling.
Since $(t_{2}-t_{1})/\t_{l}>>1$ the integral over $\z$ in Eq.(\ref{int}) will be peaked around $\z=0$, and we can therefore extend the limits of integration to $(-\infty,\infty)$. This integral and the one over $x_{2}$ may then be carried out and we obtain
\beq
\Delta=\int_{-\infty}^{0}dX\int_{-\infty}^{\infty}dp \frac{1}{2}{\rm Erfc}\left[-\sqrt{\frac{3m^{2}}{D (t_{2}-t_{1})^{3}}}\left(X+\frac{p(t_{2}-t_{1})}{m}\right)\right]W_{t_{1}}(X,p),
\eeq
where ${\rm Erfc}$ is the complementary error function \cite{Amb}. 
For large positive values of the argument we have that,
\beq
{\rm Erfc}[u]\approx \frac{e^{-u^{2}}}{u\sqrt{\pi}}\left(1+O\left(\frac{1}{u}\right)\right).
\eeq
Therefore, since the Wigner function is peaked around $p<0$, $\Delta$ will be very small, provided $(t_{2}-t_{1})<<\t_{s}$. This gives us an {\it upper} bound on the time interval, $[t_{1},t_{2}]$, rather than a lower one. The lower time scale is provided by the condition $t_{2}-t_{1}>>\t_{l}=\sqrt{2m\hbar/D}$. This lower time scale is compatible with the condition that the current be positive.

Note however that this lower limit is state independent. There will be states for which arrival time probabilities decohere on much shorter time scales than this, for example the simple cases which decohere in the absence of any environment will continue to do so in the presence of an environment, at least until a time $\sim \t_{s}$.

The key point is that whether or not one can assign arrival time probabilities in decoherent histories depends on the form of the state at the time it crosses the origin.  Environmentally induced decoherence produces mixtures of gaussian states from generic initial ones, and thus after a short time arrival time probabilities can be defined whatever the initial state. Crucially however it is not necessary for the system to be monitored whilst it is crossing the origin. The smallest time interval,  $\delta t$, over which we can define a decoherent arrival time probability is therefore set by the energy of the system and not the details of the environment and we must have $E \delta t>>\hbar$. This is in agreement with Ref.\cite{HaYe1}, and also with the results of earlier works, concerning the accuracy with which a quantum system may be used as a clock \cite{clock}.

In conclusion then, for a general state, decoherence of histories requires that we evolve for a time much greater than $\t_{l}=\sqrt{2m\hbar/D}$ before the first crossing time. This is because this is the time scale on which quantum correlations disappear and our initial state begins to resemble a mixture of gaussian states. After this time, we may define arrival time probabilities to an accuracy $\delta t$, provided only that $E\delta t>>\hbar$. States which start as gaussians may be assigned arrival time probabilities without this initial period of evolution. This is in line with the general result that some coarse-graining is always required to achieve a decoherent set of histories in quantum theory.

\section{Summary and discussion}\label{sec9}

In this paper we have been concerned with deriving an arrival time distribution for open quantum systems, and comparing this with the classical result. We began by discussing the generalisation of the current, which is the classical arrival time distribution, to open quantum systems and in particular to the case of quantum Brownian motion. We found that in general the inclusion of an environment leads to extra terms appearing in the expressions for the current, compared with the those valid in the unitary case. However we have shown that in the limit of negligible dissipation these correction terms may be dropped. We then explored the resulting arrival time distribution and showed that it is non-negative after a time of order $\t_{l}$, and that after this time it can be written as the trace of the density matrix times a POVM.

We then turned to the question of deriving this arrival time distribution from the decoherent histories approach to quantum theory.  We extended the decoherent histories analysis of the arrival time problem to the case of a particle coupled to an environment. As expected, the inclusion of an environment produces decoherence of arrival time probabilities for a generic initial state. There are, however, some limitations to the permitted class of histories. For a generic state arrival times can only be specified after an initial time $t>>\t_{l}$. Even after this time arrival times cannot be specified with arbitrary precision, coarse graining over intervals $\delta t>>\hbar/E$ is required to ensure decoherence. 
We showed that the decoherence condition is very closely related to a semi-classical approximation for the evolution of the state, and that both conditions are satisfied if the time between projections is sufficiently large. This is a specific case of a more general connection between decoherence and classical behavior.

Our approach has proceeded at two levels. At the heuristic level the simple generalisation of Eq.(\ref{1w}) to the case of a particle coupled to an environment, Eq.(\ref{curj}), is a positive arrival time distribution after a time of order $\t_{l}$. This can also be written as the expectation value of a POVM, Eq.(\ref{opcur}), and thus has the same form as other probabilities in quantum theory. On a more fundamental level, these expressions can be {\it derived} from the decoherent histories approach to quantum theory, where they are seen to be valid for times much later than $\t_{l}$. 

Although the arrival time probabilities computed from decoherent histories agree with the heuristic ones when we have decoherence, their derivation in this way represents a significant advance in our understanding. The great difficulty with regarding the current as the arrival time distribution is the arbitrary way in which one accepts these ``probabilities'' when they are positive, but declines to do so when they are not. Because decoherence is an essential part of the histories formalism, this arbitrariness is replaced with a consistent set of rules governing when probabilities may or may not be assigned. Whilst this may be of no consequence in the setting of a laboratory, it may prove hugely important in the analysis of closed systems, in particular the study of quantum cosmology \cite{jjhqc}.   

Another very interesting result we have presented is that the current becomes strictly positive after a finite time whilst assignment of probabilities in decoherent histories is only possible asymptotically. In some ways this difference between the heuristic analysis and decoherent histories is to be expected. Indeed, the current represents a linearly positive history \cite{Gold}, and it is essentially the condition of linear positivity that we have proven holds after a time of order $\t_{l}$.  It is known that linear positivity is a weaker condition than decoherence \cite{Harlp, Haqp}, and thus it is not surprising that demanding decoherence leads to a stricter limit on the assignment of probabilities than the heuristic analysis. It would be interesting to examine what is gained in this context by imposing decoherence rather than linear positivity. 

There are similarities here with the relationship between the current and Kijowski's arrival time distribution \cite{Kij}, which may be shown to agree in the classical limit, but not more generally. Indeed because it is manifestly positive, one might expect the arrival time distribution computed from decoherent histories to be more closely related to Kijowski's distribution on shorter time scales. These issues will be discussed elsewhere.

\begin{acknowledgements}
The author would like to express his thanks to J.J.Halliwell for many useful discussions, and for suggesting the problem.
\end{acknowledgements}

\bibliography{apssamp}

\end{document}